\documentclass[aps,10pt]{revtex4} 

\usepackage{amsmath,amssymb,bm,epsfig}
\usepackage{color}
\usepackage{natbib}
\usepackage{hyperref} 
\usepackage{ulem}
\usepackage{graphicx}
\usepackage{xcolor,colortbl}
\usepackage{pifont}

\newcommand{\be}{\begin{equation}}
\newcommand{\ee}{\end{equation}}
\newcommand{\ba}{\begin{eqnarray}}
\newcommand{\ea}{\end{eqnarray}}
\newcommand{\nn}{\nonumber}
\newcommand{\de}{\delta}

\newcommand{\FB}[1]{\left(#1\right)}
\newcommand{\SB}[1]{\left\{#1\right\}}
\newcommand{\TB}[1]{\left[#1\right]}

\begin{document}
\title{Medium effects on the electrical conductivity of a hot pion gas}
\author{Snigdha Ghosh$^{a,b}$}
\email{snigdha.physics@gmail.com}
\author{Sukanya Mitra$^{c}$}
\email{sukanya.mitra10@gmail.com}
\author{Sourav Sarkar$^{a,b}$}
\email{sourav@vecc.gov.in}
\affiliation{$^a$Variable Energy Cyclotron Centre, 1/AF Bidhannagar, Kolkata - 700064, India}
\affiliation{$^b$Homi Bhabha National Institute, Training School Complex, Anushaktinagar, Mumbai - 400085, India}
\affiliation{$^c$Indian Institute of Technology Gandhinagar,  Gandhinagar-382355, Gujarat, India}

\begin{abstract}

The electromagnetic response of an interacting system of pions has been studied at finite temperature.
The corresponding transport parameter {\it i.e.} electrical conductivity has been estimated by
solving the relativistic transport equation in presence of a finite electric field employing the Chapman-Enskog technique where
the collision term has been treated in the relaxation time approximation. The scattering amplitudes of charged pions
modeled by $\rho$ and $\sigma$ meson exchange using an effective Lagrangian have
been obtained at finite temperature by introducing self-energy corrections in the thermal propagators in the real time formalism. The temperature behavior of electrical conductivity for a hot pion gas shows significant quantitative 
enhancement due to the inclusion of medium effects in scattering. 

\end{abstract}
\maketitle

\section{Introduction}

The theoretical framework of strong interactions-Quantum Chromodynamics (QCD), indicates the existence of a liberated  
state made up of hadronic substructures by virtue of two crucial phenomena - color confinement and asymptotic freedom. 
In the last few decades  experimental facilities performing collisions of heavy ions at highly relativistic
energies such as the Relativistic Heavy Ion Collider (RHIC) at BNL and Large hadron Collider (LHC) at CERN have offered 
the unique opportunity to realize such a thermalized state of deconfined quarks and gluons.
After creation the system is observed to show rapid thermalization~\cite{Heinz} leading to the creation of quark-gluon
plasma (QGP). On reduction of its energy density by radial expansion it subsequently converts into a state composed of confined hadronic 
states presumably retaining considerable strong interaction.  In addition, the 
event by event proton number fluctuations in the symmetric collisions and net proton number difference in asymmetric collisions 
result in the onset of an extremely strong electromagnetic (EM) field ($B_{\pi}\sim m_{\pi}^2\sim 10^{18}$ Gauss)~\cite{Skokov,Zakharov,Toneev}, 
that has ever existed in nature. Despite being short lived, such a strong EM field casts profound effects on the dynamics of 
the created system which finally reflects on the extracted signals such as charged hadronic spectra and flow harmonics. A number of recent works~\cite{Pang,Greif,Tuchin1,Pu,Tuchin2} have studied the impact of EM field on the system produced in relativistic heavy ion collisions. In response to the strong EM field an induced current is generated within the electromagnetically charged QGP and hot hadronic matter, which is related to the 
corresponding electric field via the quantity called electric conductivity $\sigma_{el}$. It thus provides a quantitative measure
of charge transport and appears as a crucial signature of the response of the system to the created EM field.

The significance of electrical conductivity in identifying the EM responses may be observed in a number of ways.
First, the sensitivity of the charge dependent directed flow of final state hadrons on the early stage charge 
asymmetry of the created fireball, is reflected through the system's electric conductivity~\cite{Hirano,Voloshin}. 
This reveals that the charge separation effect on the measured distribution of final state particles have non-trivial
dependencies on $\sigma_{el}$. Secondly, the EM response plays a crucial role in determining the thermal photon
and dilepton emission rates entering through the current-current correlator. As a consequence $\sigma_{el}$ turns
out to be an integral input to the soft photon and dilepton emission rates~\cite{Yin,Ding,Huot}, such that their
transverse momentum spectra and elliptic flow are also sensitive on the temperature dependence of $\sigma_{el}$.
Again, the strong EM fields also affect the hydrodynamic evolution and flow of the system by enhancing the azimuthal
anisotropy of the produced particles in the overlap zone~\cite{Kharzeev,Tuchin}. In addition, $\sigma_{el}$ appears
as an initial input to the evolution equations which also influences the extracted signals by a 
considerable amount.  Hence, the precise estimation of $\sigma_{el}$ along with  
it's temperature dependence is extremely necessary for the characterization of systems under EM fields which makes electrical conductivity a topic of interest 
in the recent literature.

In the last few years, a number of estimations of the electrical conductivity were made mostly in the strongly coupled QGP
phase employing different techniques such as relativistic transport theory, dynamical quasiparticle model (DQPM), 
maximum entropy method (MEM), parton cascade model, quasiparticle approaches and so on 
\cite{Greco,Greiner-econd,Greco-econd,Cassing1,Cassing2,Qin,Patra,Mitra-Chandra}. Lattice QCD computations have
produced quite a considerable number of estimations of $\sigma_{el}$ as well~\cite{Amato,Aarts1,Aarts2,Gupta1,Brandt1,Brandt2,Ding,Francis}.
Also, a number of holographic estimations have been made for electrical conductivity 
\cite{Finazzo,Huot,Sachin-cond1,Sachin-cond2,Sachin-cond3,Bu,Rougemont1,Rougemont2}. 
In comparison to the hot QCD sector the estimation of $\sigma_{el}$  in hot hadronic
matter has received much less attention. There are nevertheless a few existing works in the hadronic sector in recent times mostly employing the correlator 
technique to obtain $\sigma_{el}$~\cite{Fraile1,Fraile2,Zahed}. In~\cite{Denicol} however, electric conductivity of a 
hadron gas has been reported upon in the relativistic kinetic theory approach using constant as well as experimental cross 
sections that include Breit-Wigner resonances. 

In the current work we follow the covariant kinetic approach to obtain $\sigma_{el}$ for a hot 
pion gas by solving the relativistic transport equation in Chapman-Enskog (CE) technique. The collision term has been
expressed in terms of thermal relaxation times of different pionic components (with different electric charges).
Finally, the dynamical input for the charge transport processes giving rise to electrical conductivity, i.e, the interaction
cross sections between different pion components, have been constructed including the relevant thermal effects of a
strongly interacting medium. This is in fact the novel feature of the study presented in this work. While all the evaluations 
we came across have used emperical $\pi\pi$ scattering cross-sections extracted from experimental cross-sections in vacuum, we have constructed a 
theoretical framework which reproduces the vacuum cross-section and is amenable to the incorporation of medium effects. Here, the scattering is taken to proceed via $\rho$ and $\sigma$ meson exchange so that medium effects enter through the self-energy corrected exact $\rho$ and 
$\sigma$ propagators evaluated at finite temperature
using the real time formalism of thermal  field theory. This technique has been earlier used to obtain the temperature dependence of viscosities~\cite{Mitra1}, thermal conductivity~\cite{Mitra2} and the relaxation time of dissipative flows~\cite{Mitra3}.  We thus obtain a realistic estimation of the 
temperature dependence of $\sigma_{el}$ 
for an interacting pionic medium at finite temperature likely to be produced in the later stages of heavy ion collisions.

The article is organized as follows. In section II the complete formalism involved in the theoretical description is given
in three consecutive subsections. The first one deals with the derivation of $\sigma_{el}$ in Chapman-Enskog method
which is followed by the estimation of thermal relaxation times of charged pion components. This section ends with
details of the interaction cross sections at finite temperature in the real time formalism. In section III
the numerical results of the temperature behavior of $\sigma_{el}$ is given along with related discussions.
We conclude the article in section IV by providing a summary and possible outlook of the present work.

\section{Formalism}

\subsection{Estimation of $\sigma_{el}$ in CE method}

We begin with the formal introduction of some essential thermodynamic quantities and their basic
definitions in a multi-component, many particle system following Ref.~\cite{Degroot}.
In order to obtain the expression of electrical conductivity for any system one compares the macroscopic and
microscopic definitions of the induced current density $J^{\mu}$.
Under the influence of an external electric field
$E^{\mu}$, the macroscopic definition of current density can be approximated by a linear relationship with the field 
itself via $\sigma_{el}$ as,
\begin{equation}
J^{\mu}=\sigma_{el} E^{\mu}~. 
\label{eq-EC1}
\end{equation}
In the microscopic definition the current density of a multi-component system is expressed in terms of the diffusion 
flow of the constituents in the following manner,
\begin{equation}
 J^{\mu}(x)=\sum_{k=1}^{N}q_{k}I_{k}^{\mu}=\sum_{k=1}^{N-1}(q_{k}-q_{N})I_{k}^{\mu}~,
 \label{eq-EC2}
\end{equation}
with $q_{k}$ as the electric charge associated with the $k^{th}$ species. The last step of Eq.~(\ref{eq-EC2}) follows
from the conservation relation $\sum_{k=1}^{N}I_{k}=0$.

The diffusion flow $I^{\mu}_a$ for a relativistic system out of equilibrium taking all reactive processes into
account, is given by,
\begin{eqnarray}
I_{a}^{\mu}=\sum_{k=1}^{N}q_{ak}I_{k}=\sum_{k=1}^{N}q_{ak}\FB{N_{k}^{\mu}-x_{k}N^{\mu}}~, 
~~~~~~~~~\FB{a=1,2,..........N'}. \label{eq-EC3}
\end{eqnarray}
Here $a$ is the index of conserved quantum number and $q_{ak}$ stands for $a^{th}$ conserved quantum number associated 
with $k^{th}$ component. $N^{\mu}(x)=\sum_{k=1}^{N}N_{k}^{\mu}(x)$ gives the expression for the total particle 4-flow, 
where, the same for the $k^{th}$ species in a multicomponent system is defined as, 
$N_k^{\mu}(x)=\int \frac{d^{3}\vec{p}_{k}}{(2\pi)^3 p_{k}^0} p_{k}^{\mu}f_{k}(x,p_{k}) $.
The particle fraction corresponding to $k^{th}$ species is defined as $x_k=\FB{\frac{n_k}{n}}$ with 
$n_k=\int \frac{d^{3}\vec{p_{k}}}{(2\pi)^3 }f_{k}(x,p_{k})$ as the particle number density of $k^{th}$ species, and 
$n=\sum_{k=1}^{N}n_{k}(x)$, as the total number density of the system respectively. Here $f_{k}(x,p_{k})$ is the single 
particle momentum distribution function for the particle $k$, which is a function of the particle four-momentum $p_k$
and space-time coordinate $x$. In an out of equilibrium situation where irreversible phenomena
 occur, the distribution function is constructed in terms of its local equilibrium value $f_{k}^{0}$ and the deviation
from it $\delta f_{k}$, so that
\begin{eqnarray}
 f_{k}=f_{k}^0 +\delta f_{k}=f_{k}^0 + f_{k}^0 \FB{1\pm f_{k}^0} \phi_{k}~.
 \label{eq-EC4}
\end{eqnarray}
The deviation function $\phi_{k}(x,p_k)$ quantifies the amount of distortion in the distribution function in an away from
equilibrium situation. It is straightforward to show that at leading order ($i.e.$ for equilibrium distribution function $f_k^0$)
the diffusion flow vanishes, while at next to leading order the deviation term $\delta f_{k}=f_{k}^0(1\pm f_k^0)$, 
gives finite contribution to the diffusion flow as
\begin{equation}
I_{a}^{\mu}=\sum_{k=1}^{N}\FB{q_{ak}-x_{a}}\int \frac{d^{3}\vec{p}_{k}}{(2\pi)^3 p_{k}^0} p_{k}^{\mu} f_k^0\FB{1\pm f_k^0}\phi_k~. 
\label{eq-EC5}
\end{equation}

Now in order to extract $\sigma_{el}$  we need to know the deviation function $\phi_k$. For this purpose we need to
solve the relativistic transport equation, i.e, the evolution equation satisfied by the particle distribution function.
In the presence of an external electromagnetic force, the relativistic transport equation for a $N$-component multi-species
system including the covariant force term is given by
\begin{equation}
 p^{\mu}_{k}\partial_{\mu}f_{k}+q_{k}F^{\alpha\beta}p_{\beta}\frac{\partial f_{k}}{\partial p_{k}^{\alpha}}
 =\sum_{l=1}^{N} C_{kl}[f_{k},f_{l}]~,~~~~~~~~ \FB{k=1,2,...N}
\label{eq-EC6}
\end{equation}
where $F^{\mu\nu}=\FB{u^\mu E^\nu-u^\nu E^\mu}$ defines the electromagnetic field tensor in the absence of any magnetic 
field and $u^{\mu}$ is the hydrodynamic four-velocity. We use the metric 
$g^{\mu\nu}=\mathrm{diag}\FB{1,-1,-1,-1}$.
The quantity on the right hand side of Eq.~(\ref{eq-EC6}) is termed as the collision term that quantifies the rate of change 
of $f_k$. Here for each $l$, $C_{kl}$ gives the collision contribution due to the scattering of $k^{th}$ particle with $l^{th}$ 
one and is defined as~\cite{AMY1},
\begin{eqnarray}
C_{kl}[f_{k},f_{l}] &=&  \FB{\frac{\nu_{l}}{1+\de_{kl}}} \frac{1}{2} \int\int\int d\Gamma_{p^{}_{l}} d\Gamma_{p'_{k}} d\Gamma_{p'_{l}} (2\pi)^4 \times\delta^{4}(p_{k}+p_{l}-p'_{k}-p'_{l}) \nn \\
&& \hspace{3cm} \times~ \langle|M_{k+l\rightarrow k+l}|^{2}\rangle \left[ \frac{}{}f_{k}(p'_{k}) f_{l}(p'_{l}) \{1\pm f_{k}(p_{k})\}\{1\pm f_{l}(p_{l})\} \right. \nn \\
&& \hspace{3.5cm}\left. -~f_{k}(p_{k})f_{l}(p_{l})\{1\pm f_{k}(p'_{k})\}\{1\pm f_{l}(p'_{l})\} \frac{}{}\right]~. \label{eq-EC7}        
\end{eqnarray}

The phase space factor is denoted by $d\Gamma_{p_{i}}=\frac{d^3 \vec {p}_{i} }{(2\pi)^3 2p^{0}_{i}}$
where $p^{0}_{i}$ is the energy of the particle. The primed notation stands
for the final state momenta corresponding to each species.
The overall $\frac{1}{2}$ factor appears due to the symmetry in order to compensate for the double counting of
final states that occurs by interchanging $p'_{k}$ and $p'_{l}$ and $\nu_{l}$ denotes the degeneracy of the scatterer $l$.  

In the present analysis we treat the collision term in relaxation time approximation (RTA) where it is assumed that all
particles except the one under study ($i.e.$ the $k^{th}$ particle) is in equilibrium.
The collision term is then expressed as the deviation of the distribution function over the thermal relaxation time 
$\tau_k$ which is actually a measure of the time scale for restoration of the out of equilibrium distribution
to its local equilibrium value. Thus,
\begin{equation}
C_{kl}[f_{k},f_{l}]=-e_{k} \frac{\delta f_{k}}{\tau_k}=-e_k \frac{f_{k}^0 (1\pm f_{k}^0) \phi_{k}}{\tau_k}~,
\label{eq-EC8}
\end{equation}
with $e_k\equiv p_{k}^0$ denoting the energy of the $k^{th}$ particle. Here $\tau_k$ has been taken as the inverse
of the reaction rate $(R_k)$ of the $k^{th}$ particle and thus is a function of its four-momentum $p_k$. Note that there
exists other ways of obtaining the relaxation time. Transport relaxation rates can be used besides other possible parametrizations. Moreover, though
RTA offers a simple and reasonably accurate way to handle the collision kernel, for better precision the 12-dimensional
collision integral given by Eq.~(\ref{eq-EC7}) needs to be considered along with more advanced methods of simplification.

We now proceed to treat the left hand side of transport equation (\ref{eq-EC6}) employing Chapman-Enskog (CE)
technique~\cite{Degroot}. It is an iterative method, where from the known lower order distribution function 
($f_{k}^{0}$) the unknown next order correction ($\phi_k$) can be determined by successive approximation. Some standard algebra
leads us to the linearized transport equation,

\begin{equation}
p_{k}^{\mu}\partial_{\mu}f_{k}^{0}+\frac{1}{T}f_{k}^0(1\pm f_{k}^0)q_{k}E_{\mu}p_{k}^{\mu}=
-\frac{e_{k}}{\tau_{k}}f_{k}^0(1\pm f_{k}^0)\phi_{k}~.
\label{eq-EC9}
\end{equation}
where the local equilibrium
distribution function for pions in terms of particle four-momenta $p_{k}^{\mu}$ and space time coordinate
$x$ is given by
\begin{equation}
 f_{k}=\frac{1}{\exp\SB{\frac{p_{k}^{\mu}u_{\mu}(x)-\mu_{k}(x)}{T(x)}}-1}~,
 \label{eq-EC10}
\end{equation}
with $\nu_{k}$ the pion chemical potential introduced for pion number conservation. The subscript $k$ here
indicates different charge states of pions. 

In order to perform the derivative on the first term of the left hand side of ($\ref{eq-EC9}$), we decompose 
the partial derivative $\partial^{\mu}$ over the distribution function into a timelike and a spacelike part 
as $\partial^{\mu}=u^{\mu}D+\nabla^{\mu}$, with $D=u^{\mu}\partial_{\mu}$ as the covariant time derivative 
and $\nabla_{\mu}=\Delta_{\mu\nu}\partial^{\nu}$ as the spatial gradient where $\Delta^{\mu\nu}\equiv g^{\mu\nu}-u^{\mu}u^{\nu}$. 
Utilizing the definition of equilibrium 
pion distribution function from (\ref{eq-EC10}) we arrive at a number of terms containing spatial
gradients and time derivatives over the thermodynamic state parameters. The spatial gradients over velocity, 
temperature, and chemical potentials can be related to the thermodynamic forces concerning the viscous flow, 
heat flow and the diffusion flow of the fluid respectively. The remaining time derivatives therefore need 
to be eliminated using a number of thermodynamic identities so that they also contribute in the expressions of
the thermodynamic forces. We list below the thermodynamic identities which are nothing but the time evolution
equations of basic thermodynamic quantities namely particle number density, energy per particle and hydrodynamic
velocity of fluid,
\begin{eqnarray}
Dn_{k} &=& -n_{k}\FB{\partial\cdot u} ~,
\label{eq-EC11}
\\
\sum_{k=1}^{N}x_{k}D\omega_{k} &=& -\frac{\sum_{k=1}^N P_{k}}{\sum_{k=1}^N n_{k}}\FB{\partial \cdot u} ~,
\label{eq-EC12}
\\
Du^{\mu} &=& \frac{\nabla^{\mu} P}{\sum_{k=1}^{N} n_{k} h_{k}}+\frac{\sum_{k=1}^N q_{k}n_{k}}{\sum_{k=1}^N h_{k}n_{k}}E^{\mu}~.
\label{eq-EC13}
\end{eqnarray}
Here $P_k$ is the partial pressure and $h_k=\FB{e_k+\frac{P_k}{n_k}}$ is the enthalpy per particle assigned to the $k^{th}$ species.
Eq.~(\ref{eq-EC13}) reveals that even though the pressure gradient is zero, the Lorentz force acting on the particles due the electric field
$E^{\mu}$ produces a non-zero acceleration. Reducing the time derivatives on the left hand side of (\ref{eq-EC9}) applying
Eq.~(\ref{eq-EC11})-(\ref{eq-EC13}), we are left with the thermodynamic forces on the left hand side of transport equation,
\begin{equation}
 \TB{\langle p_{k}^{\nu}\rangle\SB{(p_{k}.u)-h_{k}\frac{}{}}X_{qk}+\langle p_{k}^{\nu}\rangle\sum_{a=1}^{N'-1}(q_{ak}-x_{a})X_{a\nu}}=
 -\frac{T\omega_{k}}{\tau_{k}}\phi_{k}~,
\label{eq-EC14}
\end{equation}
with $X_{q\mu}$ and $X_{a\mu}$ denoting the thermal and diffusion driving forces in presence of an electric field, 
\begin{eqnarray}
X_{q\mu}&=&\TB{\frac{\partial_{\mu}T}{T}-\frac{\partial_{\mu}P}{nh}}+\TB{-\frac{1}{h}\sum_{k=1}^{N}x_{k}q_{k}E_{\mu}}~,
\label{eq-EC15}\\
X_{k\mu}&=&\TB{(\partial_{\mu}\mu_{a})_{P,T}-\frac{h_{k}}{nh}\partial_{\mu}P}+
         \TB{q_{k}-q_{N}-\frac{h_{k}-h_{N}}{h}\sum_{l=1}^{N}x_{l}q_{l}}E_{\mu}~.
\label{eq-EC16}
\end{eqnarray}
Here $h$ is the enthalpy density for the total system and $\FB{\partial_{\mu}\mu_{a}}_{P,T}=
\sum_{b=1}^{N'-1}\FB{\frac{\partial \mu_a}{\partial x_b}}_{P,T,\{x_{a}\}} \partial_{\mu}x_b$, with $x_a$ and $\mu_a$ 
being the particle fraction and chemical potential associated with $a^{th}$ quantum number respectively.
We have ignored terms related to shear and bulk viscosities since we are interested in conduction processes only.
Observing Eq.~(\ref{eq-EC15}) and (\ref{eq-EC16}), we can conclude that the terms proportional to electric field $E^{\mu}$
result from the EM response within the medium and will contribute only to electrical conductivity. The structure of the 
left hand side of transport equation leads us to construct the deviation function $\phi_{k}$ on the right hand side 
as a linear combination of the thermodynamic forces as,
\begin{equation}
\phi_{k}=B_{k}^{\mu}X_{q\mu}+\frac{1}{T}\sum_{a=1}^{N'-1}B_{ak}^{\mu}X_{a\mu}~,
\label{eq-EC17}
\end{equation}
with $B_{k}^{\mu}$ and $B_{ak}^{\mu}$ the unknown coefficients which need to be determined. They can be obtained by
putting Eq.~(\ref{eq-EC17}) on the right hand side of Eq.~(\ref{eq-EC14}) and comparing both the sides. Utilizing the
fact that thermodynamic forces are independent these coefficients come out to be,
\begin{eqnarray}
B_{k}^{\mu}&=& \langle p_{k}^{\mu}\rangle\TB{\frac{\omega_{k}-h_{k}}{\FB{-\frac{T\omega_{k}}{\tau_k}}}}~,
\label{eq-EC18}\\
B_{ak}^{\mu}&=& T\langle p_{k}^{\mu}\rangle\TB{\frac{q_{ak}-x_{a}}{\FB{-\frac{T\omega_{k}}{\tau_k }}}}~.
\label{eq-EC19}
\end{eqnarray}

Having determined the complete structure of $\phi_k$ in terms of thermal relaxation times of constituents we replace it
on the right hand side of (\ref{eq-EC5}) to obtain the linear law of diffusion flow as,
\begin{equation}
I_{a}^{\mu}=l_{aq}X_{q}^{\mu}+\sum_{b=1}^{N'-1}l_{ab}X_{b}^{\mu}~,~~~~~~~~~~ a=1,....,(N'-1)~,
\label{eq-EC20}
\end{equation}
where the coefficients associated with thermal diffusion and particle concentration diffusion are respectively given by,
\begin{eqnarray}
l_{aq} &=&\sum_{k=1}^{N}\FB{q_{ak}-x_{a}} \int \frac{d^{3}\vec{p}_{k}}{(2\pi)^3}f_k^0(1\pm f_k^0)\frac{\tau_{k}}{T}\frac{|\vec{p}_k|^2}{e_k^2}(e_k-h_k),\nonumber
\label{eq-EC21}\\
l_{ab} &=&
\sum_{k=1}^{N}(q_{ak}-x_{a})(q_{bk}-x_{b})\int \frac{d^{3}\vec{p}_{k}}{(2\pi)^3}f_k^0(1\pm f_k^0)\frac{\tau_{k}}{T}\frac{|\vec{p}_k|^2}{e_k^2}.\nonumber
\label{eq-EC22}
\end{eqnarray}

Finally, substituting the expression of diffusion flow from Eq.~(\ref{eq-EC20}) into the microscopic definition of current density 
in Eq.~(\ref{eq-EC2}), and keeping the terms proportional to electric field only we finally obtain the expression for the 
electric current density as,
\begin{eqnarray}
 J^{\mu}=\sum_{k=1}^{N-1}(q_k-q_N)\TB{\sum_{l=1}^{N-1}l_{kl}\SB{q_l-q_N-
\frac{h_l-h_N}{h}\sum_{n=1}^{N}x_{n}q_{n}}
 -\frac{l_{kq}}{h}\sum_{n=1}^{N}x_{n}q_{n}}E^{\mu}~.
 \label{eq-EC23}
\end{eqnarray}

Comparing Eq.~(\ref{eq-EC23}) with the macroscopic definition of induced current density given by Eq.~(\ref{eq-EC1})
we arrive at the expression for electrical conductivity for an $N$ component system as \cite{Degroot},
\begin{eqnarray}
 \sigma_{el}=\sum_{k=1}^{N-1}(q_k-q_N)\TB{\sum_{l=1}^{N-1}l_{kl}\SB{q_l-q_N-
\frac{h_l-h_N}{h}\sum_{n=1}^{N}x_{n}q_{n}}
 -\frac{l_{kq}}{h}\sum_{n=1}^{N}x_{n}q_{n}}~.
 \label{eq-EC24}
\end{eqnarray}

For a three component ($\pi^{+}, \pi^{-}$ and $\pi^{0}$)  system of pions the expression of electrical conductivity
reduces to,
\begin{eqnarray}
 \sigma_{el}&=& e^2\left[(l_{11}+l_{21})\SB{1-\FB{\frac{h_{1}-h_{3}}{h}}(x_1+x_2)}
                     +(l_{12}+l_{22})\SB{1-\FB{\frac{h_{2}-h_{3}}{h}}(x_1+x_2)} \right. \nonumber \\
             && \hspace{3cm} \left. -~(l_{1q}+l_{2q})\frac{(x_1+x_2)}{h}\right]  ~,       
\label{eq-EC25}
\end{eqnarray}
where the electronic charge is given in terms of the fine structure constant $\frac{e^2}{4\pi}=\FB{\frac{1}{137}}$.

\subsection{Thermal relaxation times of pion components}

We now specify the thermal relaxation times for separate pionic charge states. 
The expression of the relaxation times for different species in a multicomponent system can be obtained by
putting Eq.~(\ref{eq-EC4}) into the right hand side of Eq.~(\ref{eq-EC7}) and assuming that all except the $k^{th}$ particle are in 
equilibrium. Comparing with Eq.~(\ref{eq-EC8}), the relaxation time is obtained as the inverse of the reaction rate $R_k$~\cite{Zhang},
\begin{eqnarray}
\tau_{k}^{-1}(p_k)\equiv R_{k}(p_k) &=& \sum_{l=1}^{N}\FB{\frac{\nu_{l}}{1+\de_{kl}}} \frac{1}{2e_{k}} \int\int\int d\Gamma_{p_{l}} d\Gamma_{p'_{k}} d\Gamma_{p'_{l}} \FB{2\pi}^4
\delta^{4}(p_{k}+p_{l}-p'_{k}-p'_{l}) \nn \\
&& \times~|\mathcal{M}_{k+l\rightarrow k+l}|^{2} \times \frac{f_{l}^{0} (1\pm f_{k}^{'0}) (1\pm f_{l}^{'0})}{(1\pm f_{k}^{0})}~, ~~~~~~~~~\FB{k=1,\cdots ,N}~.
\label{eq-EC26}
\end{eqnarray}
Here, $\mathcal{M}_{k+l\rightarrow k+l}$ is the amplitude for binary 
elastic scattering processes involving charged pion states, to be specified in the next section. Considering now that 
the momentum transfer $q=|\vec{p_k}-\vec{p'_{k}}|=|\vec{p_{l}}-\vec{p'_{l}}|$ is not too large, the following assumptions 
can be made, $f_{k}^{0}\cong f_{k}^{'0}$ and $f_{l}^{0}\cong f_{l}^{'0}$ ~\cite{Thoma}.

In centre of momentum (CM) frame the expression of $\tau_k$ reduces to,

\begin{eqnarray}
\tau_{k}^{-1}(p_k) &=& \sum_{l=1}^{N}\frac{1}{32\pi e_k}\FB{\frac{\nu_{l}}{1+\de_{kl}}}\int d\Gamma_l \frac{\lambda^{1/2}\FB{s,m_k^2,m_l^2}}{s} d(\cos\theta_{CM}) \nn \\
&& \hspace{4cm} \times f_{l}^0(1\pm f_{l}^0)|\mathcal{M}_{k+l\rightarrow k+l}|^2, ~~~~~~~~ ~\FB{k=1,\cdots ,N}~, \nonumber
\end{eqnarray}
where $s$ is the CM energy and $\lambda(x,y,z)=x^2+y^2+z^2-2xy-2yz-2zx$ is the triangular function.
$\theta_{CM}$ is the scattering angle in the CM frame. 
%
%
\subsection{Estimation of $\pi\pi$ cross section at finite temperature}
\begin{figure}[h]
\begin{center}
\includegraphics[angle=-90, scale=0.5]{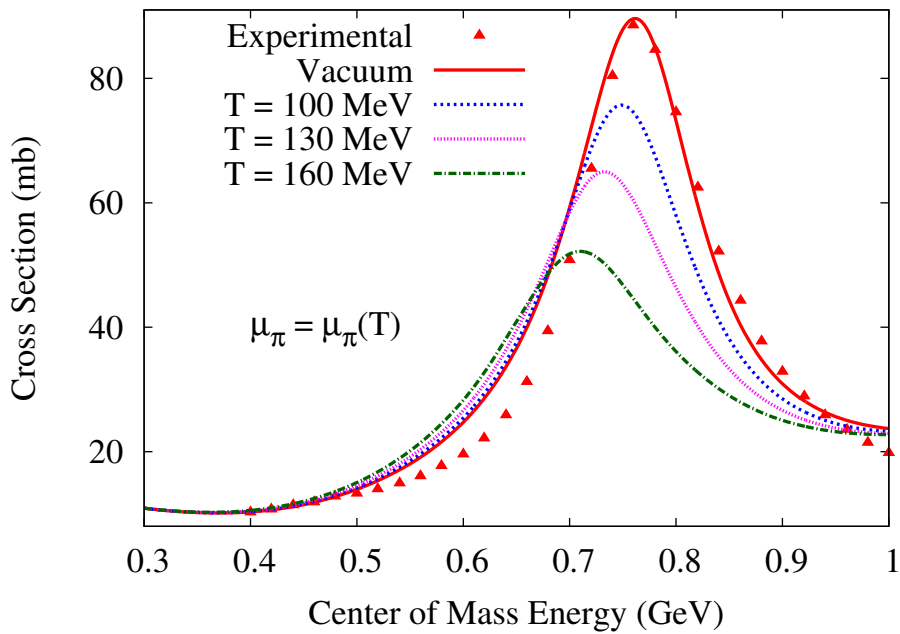}
\end{center}
\caption{The isospin averaged total in-medium cross sections of elastic $\pi\pi\rightarrow\pi\pi$ scattering at three different
 temperatures (100, 130 and 160 MeV respectively) compared with vacuum cross section and experimental data~\cite{Prakash:1993bt}. 
 In all the cases, temperature dependent pion chemical potential has been used.}
\label{fig.xsection}
\end{figure}
To calculate the Lorentz invariant amplitudes for elastic binary scattering among different charge states of pions, we
take the following well known effective $\rho\pi\pi$ and $\sigma\pi\pi$ interations~\cite{Serot:1984ey},
\begin{eqnarray}
\mathcal{L}_{\rho\pi\pi} &=& -g_{\rho\pi\pi}\vec{\rho}_\mu\cdot \left( \vec{\pi}\times\partial^\mu\vec{\pi} \right) \label{eq.lagrangian_rhopipi}\\
\mathcal{L}_{\sigma\pi\pi} &=& g_{\sigma\pi\pi} \sigma\left(\vec{\pi}\cdot\vec{\pi}\right) \label{eq.lagrangian_sigmapipi}
\end{eqnarray} 
where, $g_{\rho\pi\pi}$ = 6.05 and $g_{\sigma\pi\pi}$ = 0.525 GeV. 

The invariant amplitudes for $\pi(k)\pi(p)\rightarrow\pi(k^\prime)\pi(p^\prime)$ scattering processes are 
\begin{eqnarray}
\mathcal{M}_{\pi^+\pi^+\rightarrow\pi^+\pi^+} &=& -\left(\mathcal{M}_t^\rho+\mathcal{M}_u^\rho\right) 
+ 4\left(\mathcal{M}_t^\sigma+\mathcal{M}_u^\sigma\right) \nonumber \\
\mathcal{M}_{\pi^+\pi^-\rightarrow\pi^+\pi^-} &=& \left(\mathcal{M}_s^\rho+\mathcal{M}_t^\rho\right) 
+ 4\left(\mathcal{M}_s^\sigma+\mathcal{M}_t^\sigma\right) \nonumber \\
\mathcal{M}_{\pi^+\pi^0\rightarrow\pi^+\pi^0} &=& \left(\mathcal{M}_s^\rho-\mathcal{M}_u^\rho\right) 
+ 4\mathcal{M}_t^\sigma \nonumber \\
\mathcal{M}_{\pi^-\pi^+\rightarrow\pi^-\pi^+} &=& \mathcal{M}_{\pi^+\pi^-\rightarrow\pi^+\pi^-} \nonumber \\
\mathcal{M}_{\pi^-\pi^-\rightarrow\pi^-\pi^-} &=& \mathcal{M}_{\pi^+\pi^+\rightarrow\pi^+\pi^+} \nonumber \\
\mathcal{M}_{\pi^-\pi^0\rightarrow\pi^-\pi^0} &=& \mathcal{M}_{\pi^+\pi^0\rightarrow\pi^+\pi^0} \nonumber \\
\mathcal{M}_{\pi^0\pi^+\rightarrow\pi^0\pi^+} &=& \mathcal{M}_{\pi^+\pi^0\rightarrow\pi^+\pi^0} \nonumber \\
\mathcal{M}_{\pi^0\pi^-\rightarrow\pi^0\pi^-} &=& \mathcal{M}_{\pi^0\pi^+\rightarrow\pi^0\pi^+} \nonumber \\ 
\mathcal{M}_{\pi^0\pi^0\rightarrow\pi^0\pi^0} &=& 4 \left( \mathcal{M}_s^\sigma + \mathcal{M}_t^\sigma 
+ \mathcal{M}_u^\sigma \right) \nonumber ~,
\end{eqnarray}
where, 
\begin{eqnarray}
\mathcal{M}_s^\rho &=& g_{\rho\pi\pi}^2 \left[ \frac{t-u}{s-m_\rho^2-\Pi_\rho} \right] \label{eq.M.s.rho} \\
\mathcal{M}_t^\rho &=& g_{\rho\pi\pi}^2 \left[ \frac{s-u}{t-m_\rho^2} \right] \label{eq.M.t.rho} \\
\mathcal{M}_u^\rho &=& g_{\rho\pi\pi}^2 \left[ \frac{s-t}{u-m_\rho^2} \right] \label{eq.M.u.rho} \\
\mathcal{M}_s^\sigma &=& g_{\sigma\pi\pi}^2 \left[ \frac{1}{s-m_\sigma^2-\Pi_\sigma} \right] \label{eq.M.s.sigma} \\
\mathcal{M}_t^\sigma &=& g_{\sigma\pi\pi}^2 \left[ \frac{1}{t-m_\sigma^2} \right] \label{eq.M.t.sigma} \\
\mathcal{M}_u^\sigma &=& g_{\sigma\pi\pi}^2 \left[ \frac{1}{u-m_\sigma^2} \right] \label{eq.M.u.sigma}.
\end{eqnarray}
Eqs.~(\ref{eq.M.s.rho})-(\ref{eq.M.u.sigma}) correspond to the contributions from different Feynman diagrams 
where, $s=\left(k+p\right)^2$, $t=\left(k-k^\prime\right)^2$ and 
$u=\left(k-p^\prime\right)^2$ are the Mandelstam variables. Note that in the $s$-channel diagrams we have used effective propagators
 for the $\rho$ and $\sigma$ obtained from a Dyson-Schwinger sum 
 in which $\Pi_\rho$ and $\Pi_\sigma$ are the self energies of $\rho$ and $\sigma$ meson respectively. 
 
Our next task is to evaluate the one-loop self energies of $\rho$ and $\sigma$ meson. For the $\rho$, we have taken 
contributions from $\pi\pi$, $\pi\omega$, $\pi h_1$ and $\pi a_1$ loop diagrams~\cite{Ghosh:2009bt}, whereas for $\sigma$
only the $\pi\pi$ loop is considered. First, we write down the expressions for the spin averaged self energies in {\it vacuum},
\begin{eqnarray}
\Pi_\rho^{vac}(q) &=& \sum\limits_{h\in\{\pi,\omega,h_1,a_1\}}^{}i\int\frac{d^4k}{(2\pi)^4}N_{\pi h}(q,k)\Delta_F(k)\Delta_F(p=q-k) \label{eq.rho.self.vac} \\
\Pi_\sigma^{vac}(q) &=& i\int\frac{d^4k}{(2\pi)^4}N_\sigma(q,k)\Delta_F(k)\Delta_F(p=q-k) \label{eq.sigma.self.vac},
\end{eqnarray} 
where, $\Delta_F(k)=\left(\frac{-1}{k^2-m_k^2+i\epsilon}\right)$ is the scalar Feynman propagator with mass $m_k$; $N_{\pi h}$ and
$N_\sigma$ contains terms coming from interaction vertices as well as from the numerator of the vector propagators in cases where $h=\omega,h_1$ or $a_1$. The explicit forms of $N_{\pi h}$ and $N_\sigma$ are~\cite{Ghosh:2009bt},
\begin{eqnarray}
N_{\pi\pi}(q,k) &=& -\left(\frac{1}{3}\right) g_{\rho\pi\pi}^2 \left[q^2+4k^2-4q\cdot k \frac{}{}\right] \nonumber \\
N_{\pi\omega}(q,k) &=& -\left(\frac{8}{3}\right) g_{\rho\pi\omega}^2 \left[k^2q^2-(q\cdot k)^2 \frac{}{}\right] \nonumber \\
N_{\pi h_1}(q,k) &=& -\left(\frac{1}{3}\right) g_{\rho\pi h_1}^2 \left[-k^2q^2-2(q\cdot k)^2 + \frac{q^2}{m_{h_1}^2}
\left( k^2q^2-(q\cdot k)^2\right)\right] \nonumber \\
N_{\pi a_1}(q,k) &=& -\left(\frac{2}{3}\right) g_{\rho\pi h_1}^2 \left[-k^2q^2-2(q\cdot k)^2 + \frac{q^2}{m_{a_1}^2}
\left( k^2q^2-(q\cdot k)^2\right)\right] \nonumber \\
N_\sigma(q,k) &=& 8 g_{\sigma\pi h_1}^2 \nonumber
\end{eqnarray}
where, $g_{\rho\pi\omega}$ = 9.35 GeV$^{-1}$, $g_{\rho\pi h_1} $ = 10.75 GeV$^{-1}$ and $g_{\rho\pi a_1}$=11.82 GeV$^{-1}$.
In order to calculate the corresponding {\it in-medium} self energies (at finite temperature and chemical potential), we have used the Real 
Time Formalism (RTF) of thermal field theory~\cite{bellac} in which all two-point functions including the 1-loop self energy become $2\times2$
 matrices. However they can be diagonalized in terms of analytic functions. Following standard prescription~\cite{bellac,Mallik:2016anp},
 we obtain the real and imaginary parts of the in-medium self energy functions and they can be written as,
\begin{eqnarray}
\text{Re}~\Pi_{\rho,\sigma}(q) &=& \text{Re}~\Pi_{\rho,\sigma}^{vac}(q) + \sum\limits_{h}^{} \int\frac{d^3k}{(2\pi)^3}\frac{1}{2\omega_k\omega_p}\mathcal{P}\left[
\left(\frac{\eta_+^k\omega_p \mathcal{N}_{\pi h,\sigma}(k^0=\omega_k)}{(q_0-\omega_k)^2-\omega_p^2}\right) \right. \nonumber \\
&& \hspace{2cm} \left.+~\left(\frac{\eta_-^k\omega_p \mathcal{N}_{\pi h,\sigma}(k^0=-\omega_k)}{(q_0+\omega_k)^2-\omega_p^2}\right) 
+\left(\frac{\eta_+^p\omega_k \mathcal{N}_{\pi h,\sigma}(k^0=q_0-\omega_p)}{(q_0-\omega_p)^2-\omega_k^2}\right) \right. \nonumber \\
&& \hspace{3cm}\left. +~\left(\frac{\eta_-^p\omega_k \mathcal{N}_{\pi h,\sigma}(k^0=q_0+\omega_p)}{(q_0+\omega_p)^2-\omega_k^2}\right) \right] \label{eq.re}
\end{eqnarray}
\begin{eqnarray}
\text{Im}~\Pi_{\rho,\sigma}(q) &=& -\pi\epsilon(q_0)\sum\limits_{h}^{}\int\frac{d^3k}{(2\pi)^3}\frac{1}{4\omega_k\omega_p} \times 
\nonumber \\ &&
\hspace{2cm}\left[\frac{}{}\mathcal{N}_{\pi h,\sigma}(k^0=\omega_k)\left\{(1+\eta_+^k+\eta_+^p)\delta(q_0-\omega_k-\omega_p) \frac{}{}\right. \right. \nonumber \\
&& \hspace{5.5cm}\left. \left. +~(-\eta_+^k+\eta_-^p) \delta(q_0-\omega_k+\omega_p)\frac{}{} \right\} \right. \nonumber  \\ 
&& \hspace{2cm}\left. +~ \mathcal{N}_{\pi h,\sigma}(k^0=-\omega_k)\left\{(-1-\eta_-^k-\eta_-^p)\delta(q_0+\omega_k+\omega_p) \frac{}{} \right.\right. \nonumber \\ 
&& \hspace{5.5cm}\left.\left. +~ (\eta_-^k-\eta_+^p)\delta(q_0+\omega_k-\omega_p)\frac{}{}\right\} \frac{}{} \right], \label{eq.im}
\end{eqnarray}
where, $\omega_k=\sqrt{\vec{k}^2+m_k^2}$, $\omega_p=\sqrt{\vec{p}^2+m_p^2} =\sqrt{(\vec{q}-\vec{k})^2+m^2_p}$, 
$\eta_\pm^k=\left[ e^{(k.u\mp\mu_k)/T}-1 \right]^{-1}$ is the Bose-Einstein distribution function with $u^\mu$ being the
4-velocity of the medium. In the local rest frame $u^\mu\equiv(1,\vec{0})$. 
To take into account the finite vacuum width of unstable particles ($h_1$ and $a_1$) in the loop, contributions to the self energy functions from $\pi h_1$ and $\pi a_1$ loops are convoluted with vacuum spectral functions of $h_1$ and $a_1$ respectively~\cite{Sarkar:2004jh}.
 
To check how the results of the model compare with experimental data, it is convenient to go to the isospin basis. The 
differential scattering cross section is given by 
$\frac{d\sigma}{d\Omega}=\left(\frac{|\bar{\mathcal{M}}|^2}{64\pi^2 s}\right)$ where, the isospin averaged invariant amplitude is
\begin{eqnarray}
|\bar{\mathcal{M}}|^2 = \frac{1}{9}\sum\limits_{I=0}^{2}\left(2I+1\right)|\mathcal{M}_I|^2 \nn
\end{eqnarray}
in which,
\begin{eqnarray}
\mathcal{M}_0 &=& 2\left(\mathcal{M}_t^\rho+\mathcal{M}_u^\rho\right)+4\left( 3\mathcal{M}_s^\sigma+\mathcal{M}_t^\sigma+\mathcal{M}_u^\sigma \right) \nn \\
\mathcal{M}_1 &=& \left(2\mathcal{M}_s^\rho+\mathcal{M}_t^\rho-\mathcal{M}_u^\rho\right)+4\left( \mathcal{M}_t^\sigma-\mathcal{M}_u^\sigma \right) \nn \\
\mathcal{M}_2 &=& -\left(\mathcal{M}_t^\rho+\mathcal{M}_u^\rho\right)+4\left( \mathcal{M}_t^\sigma+\mathcal{M}_u^\sigma \right). \nn
\end{eqnarray}

The total isospin averaged scattering cross section ($\sigma_{TOT}$) plotted against the center of mass energy ($\sqrt{s}$) is shown in Fig.~\ref{fig.xsection}.
It is seen that the vacuum cross section denoted by the solid curve 
agrees fairly well with the experimental data~\cite{Prakash:1993bt} on ignoring the non-resonant $I=2$ contribution which leads to an overestimation. The model is thus normalized to experimental data in vacuum.

A discussion of the pion chemical potential is in order at this point. In heavy ion collisions 
pions get out of chemical equilibrium early,
at $T\sim$ 170 MeV, and a corresponding chemical potential starts building up
with decrease in temperature. The kinetics of the gas is then dominated by 
elastic collisions including resonance formation such as 
$\pi\pi\leftrightarrow\rho$ etc. At still lower temperature, $T\sim$ 100 MeV
elastic collisions become rarer and the momentum distribution gets frozen
resulting in kinetic freeze-out. This scenario is quite compatible with the
treatment of medium modification of the $\pi\pi$ cross-section being employed
in this work where the $\pi\pi$ interaction is mediated by $\rho$ and $\sigma$
exchange and the subsequent propagation of these mesons are modified by two-pion
and effective multi-pion fluctuations. 
We take the temperature dependent pion chemical potential from 
Ref.~\cite{Hirano:2002ds} which implements the formalism described in~\cite{Bebie:1991ij}
and reproduces the slope of the transverse momentum spectra of identified
hadrons observed in experiments.
Here, by
fixing the ratio $s/n$ where $s$ is the entropy density and $n$ the number
density, to the value at chemical freeze-out where $\mu_\pi=0$,
one can go down in
temperature up to the kinetic freeze-out by increasing the pion chemical
potential. This provides the temperature dependence leading to $\mu_\pi(T)$
whose value starts from zero at chemical freezeout and rises to a maximum at kinetic freezeout. The temperature dependence is parametrized as
\be
\mu_\pi(T)=a+bT+cT^2+dT^3 \nn
\ee
with $a=0.824$, $b=3.04$, $c=-0.028$, $d=6.05\times 10^{-5}$ and $T$, $\mu_\pi$ in MeV.  
In this partial chemical equilibrium scenario of~\cite{Bebie:1991ij}
the chemical potentials of the heavy mesons are determined from elementary
processes. The $\omega$ chemical potential e.g. is given  by
$\mu_\omega=3\times0.88\mu_\pi$,
 as a consequence of the processes 
$\omega\leftrightarrow\pi\pi\pi$ occurring in the medium.
The branching ratios are taken from~\cite{PDG}.

The in-medium cross-section is now obtained by introducing the effective propagators for the $\rho$ and $\sigma$ mesons in the expressions for the matrix elements. As discussed above, $\omega\pi$, $h_1\pi$ and $a_1\pi$ loops modify $\rho$ 
propagation in addition to $\pi\pi$. Owing to the large $3\pi$ decay widths of the $\omega$, $h_1$ and $a_1$ these loops may be considered as multi-pion fluctuations.
In Fig.~\ref{fig.xsection}, the in-medium cross sections are shown at three different temperatures 
($T$ = 100, 130 and 160 MeV respectively) incorporating $\mu_\pi(T)$. As the temperature increases the thermal distribution functions
 in Eq.~(\ref{eq.im}) increase resulting in increase in the imaginary parts of the self energies. Physically it implies
 enhancement of decay and scattering rates of $\rho$ and $\sigma$ in the thermal medium. This ultimately 
 suppresses the in-medium cross section and this suppression increases with the increase in temperature.
  A small shift of the peak of the cross section towards lower $\sqrt{s}$ with the increase in 
 temperature is due to the small positive contribution from the real part of the self energies in Eq.~(\ref{eq.re}).

\section{Results and Discussions}
\begin{figure}
\begin{center}
\includegraphics[angle=-90, scale=0.5]{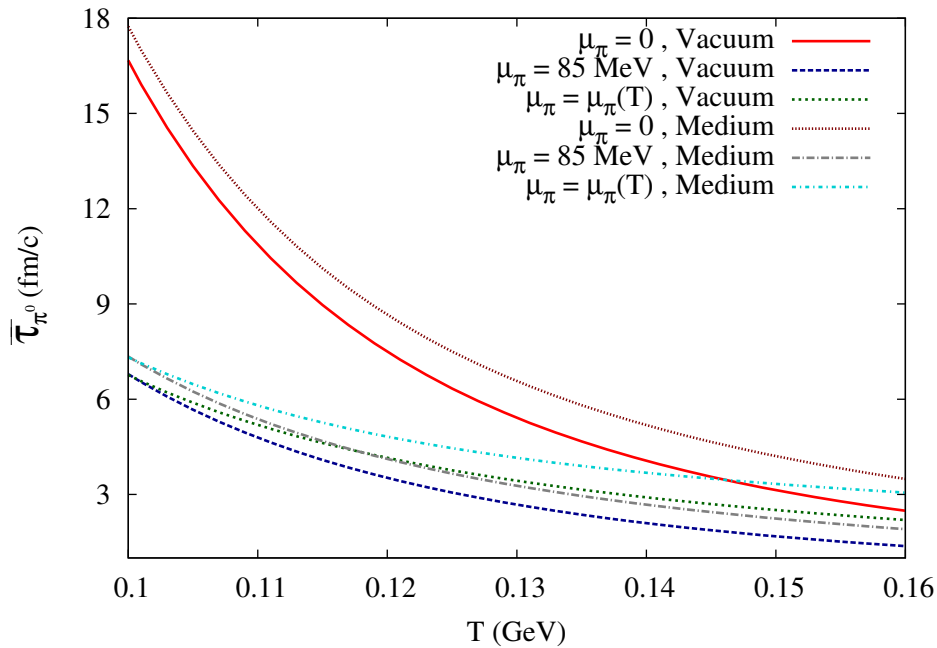} 
\end{center}
\caption{Relaxation time as a function of $T$ for $\pi^{0}$ for different cross sections and pion chemical potentials}
\label{taupi0}
\end{figure}
\begin{figure}
\begin{center}
\includegraphics[angle=-90, scale=0.5]{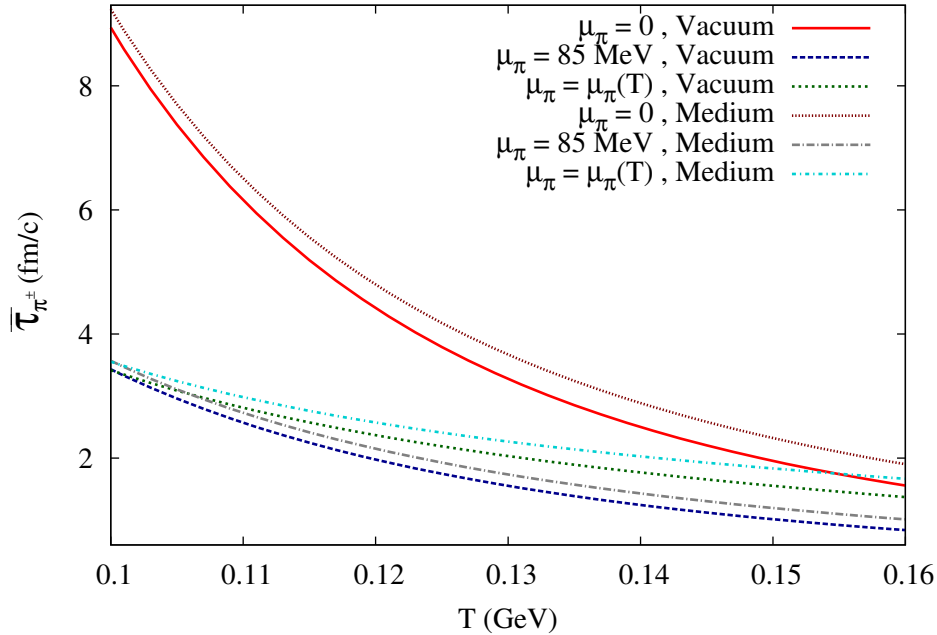}
\end{center}
\caption{Relaxation time as a function of $T$ for $\pi^{\pm}$ for different cross sections and pion chemical potentials}
\label{taupipm}
\end{figure}
We begin our discussion of numerical results with the temperature dependence of the mean relaxation times for $\pi^0$ and $\pi^\pm$.  For species $k$ this is defined in terms of the thermal average of the momentum dependent inverse relaxation time
$R_k(p_k)=1/\tau_k(p_k)$ as
\be 
\bar R_k(T,\mu)=\int d^3p_k R_k(p_k)f_k^0(p_k){\bigg/}\int d^3p_k f_k^0(p_k) \nn
\ee
so that, the mean relaxation time is $\bar\tau_k (T,\mu)=1/\bar R_k(T,\mu)$.
In Fig.~\ref{taupi0} and (\ref{taupipm}) we plot the mean relaxation time for $\pi^0$ and $\pi^\pm$ respectively as a function of $T$ for three different values of pion
chemical potentials; $\mu_{\pi}=0$, $\mu_{\pi}=85$ MeV and $\mu_{\pi}=\mu_{\pi}(T)$, the latter interpolating between the values at chemical and kinetic freezeout respectively. The features of the numerical results can be understood by realizing that the relaxation time for binary collision approximately goes as $~1/\sigma n$ where $n$ is the density and $\sigma$ is the cross-section. The increase in number density $n$ of a species with $T$ plays the dominant role and is responsible for the decreasing nature of the curves. Each set of curves correspond to vacuum and medium cross-sections for the same value of pion chemical potential. The ones corresponding to medium lie above the vacuum since the cross-section is lower in the medium as seen in Fig.~\ref{fig.xsection}. As discussed above, additional decay and scattering processes in the medium leads to a larger imaginary part and hence to a lower cross-section. Also, the amplitudes being larger, the relaxation time for the charged pions is lower compared to the neutral ones leading to a faster restoration of equilibrium for the charged components.

\begin{figure}
\begin{center}
\includegraphics[angle=-90, scale=0.5]{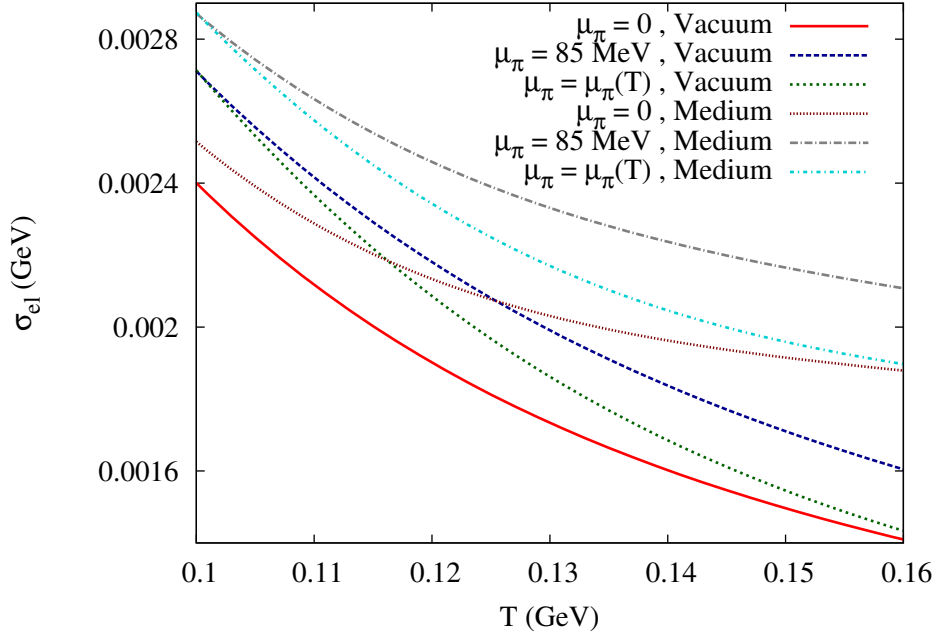} 
\end{center}
\caption{$\sigma_{el}$ as a function of $T$ for different cross sections and pion chemical potentials}
\label{sigma}
\end{figure}

\begin{figure}
\begin{center}
\includegraphics[angle=-90, scale=0.5]{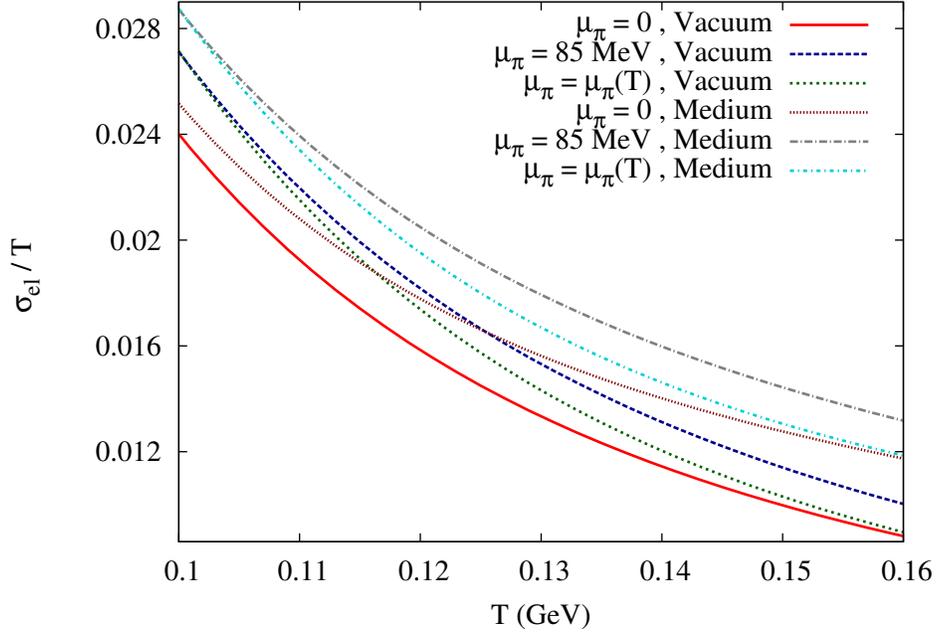}
\end{center}
\caption{$\sigma_{el}/T$ as a function of $T$ for different cross sections and pion chemical potentials}
\label{sigmabyT}
\end{figure}
Next, we discuss results for the electrical conductivity of the pion gas at finite temperature. First we show the 
estimation of $\sigma_{el}$ as a function of temperature in Fig.~\ref{sigma}. Since the relaxation times discussed above contain the dynamical input
to the electrical conductivity in the present approach, the influence  of the thermal medium as well as that of the temperature dependent pion chemical
potential gets reflected in the temperature dependence of $\sigma_{el}$. As is usually done in the literature, we also plot the dimensionless ratio $\sigma_{el}/T$
 as a function of temperature in Fig.~\ref{sigmabyT} to get an estimation of charge transport in a strongly interacting pion gas at finite
temperature. The temperature behavior shows a decreasing trend with increasing $T$, which agrees with the 
results obtained employing relativistic kinetic theory approach~\cite{Denicol} as well as using correlator
techniques~\cite{Fraile2}. For each value of $\mu_{\pi}$ the medium modified thermal $\rho$ and $\sigma$ propagators which
cause a suppression of the cross section, create visible enhancement in the $T$ dependence of $\sigma_{el}$. The larger medium
effects at higher temperatures indicate a larger broadening of $\rho$ and $\sigma$ widths with the increase of temperature. 
This shows that at finite temperature all the additional scattering and decay processes other than the normal
decay of the resonances ($\rho$ and $\sigma$) contributing to the $\pi\pi$ cross section, play an extremely significant role
in deciding the temperature behavior of $\sigma_{el}$. The plots with $\mu_{\pi}(T)$ with and without medium effects interpolate between
the plots with $\mu_{\pi}=0$ and $85$ MeV, indicating the fact that the pion number is being conserved within the temperature range
between chemical to kinetic freeze out.

\begin{figure}
\begin{center}
\includegraphics[angle=0, scale=0.5]{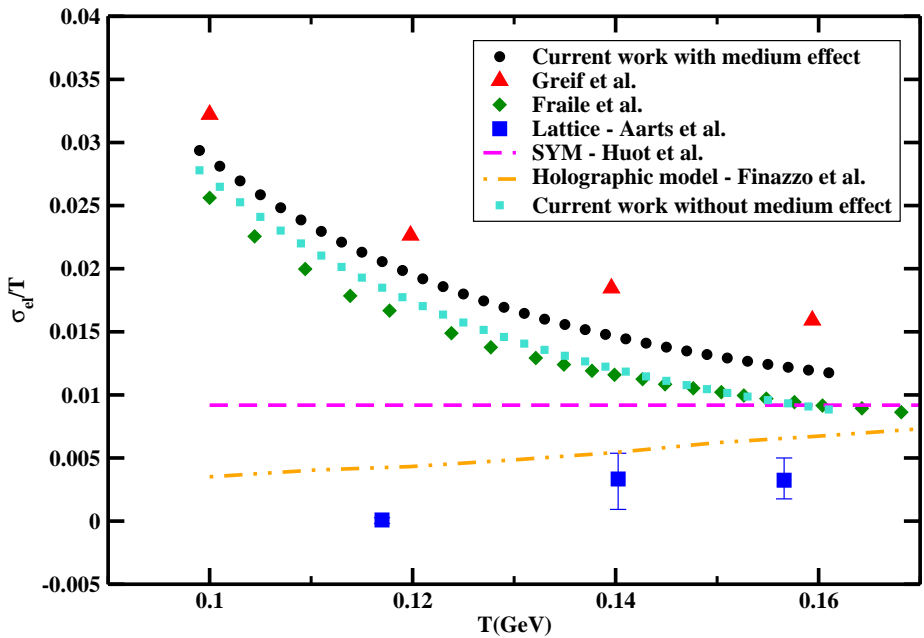} 
\end{center}
\caption{Comparison of $\sigma_{el}/T$ vs. $T$ with other estimations}
\label{sigma_compare}
\end{figure}

Finally, in Fig.~\ref{sigma_compare} we compare our results of electrical conductivity with other estimations
available in the recent literature. As observed, our estimation of $\sigma_{el}/T$ with the medium modified cross section is 
quite close to the pion gas estimation of~\cite{Denicol} where electrical conductivity has been obtained from linearized Boltzmann
equation using the $\rho$ mediated parameterized cross section. Our vacuum results are also remarkably similar to
the estimation of~\cite{Fraile2} where unitarized pion cross section has been used. This agreement can be understood by realizing that the leading order result in Kubo formalism actually gives the same result as the relaxation time approximation used in the current work and unitarization in
effect dynamically introduces the $\rho$ and $\sigma$ resonances in the $\pi\pi$ scattering amplitude that was initially absent
in the lowest order ChPT interaction. The lattice data near the transition temperature $T_c$ has been taken from~\cite{Aarts1} 
for 2+1 flavour anisotropic lattices.
There is a quantitative disagreement of our estimation with lattice results like all other pQCD estimation of $\sigma_{el}$
above $T_c$ that are $\sim$ 10 times larger than the lattice values. Also a holographic estimate of $\sigma_{el}$ from~\cite{Finazzo}
and $N=4$ super-Yang-Mills theory based calculation from~\cite{Huot} have been mentioned for comparison. These ADS/CFT based calculations
appear comparable with our results as one approaches towards $T_c$.

\section{Conclusion and outlook}

In the current article we have reported on the estimation of the electrical conductivity of an interacting pion gas at finite
temperature. The expression for electrical conductivity has been derived from the relativistic transport equation following
Chapman-Enskog technique and treating the collision term in terms of thermal relaxation times of the pionic components
$\pi^{0}$ and $\pi^{\pm}$. The mutual interactions between the charged components which actually conducts the current and
the uncharged component that acts as a resistance to the current provide the dynamical inputs for evaluating $\sigma_{el}$.
The formula for $\sigma_{el}$ given by Eq.~(\ref{eq-EC25}) has been expressed explicitly in terms of the 
respective cross sections. It is easily realized that in the non-relativistic limit it reduces to
the Drude formula for electrical conductivity $\sigma_{el}=\FB{\frac{nq^2\tau}{m}}$ with $m$ as the charge particle mass.
The key ingredient of the current work is the introduction of the effects of a thermal medium which is likely to be created in
the later stages of heavy ion collisions. After ensuring the agreement of the $\pi\pi$ interaction cross section evaluated using explicit $\rho$ and $\sigma$ meson exchange with the 
experimentally measured data, in-medium effects have been inserted using the standard techniques of thermal field theory.
The medium modified meson propagators cause a suppression of the pion cross section due to increase of their widths in the thermal bath
which is seen to affect the temperature dependence of $\sigma_{el}$ quite significantly.
We find a decreasing trend of electrical conductivity with the rise of temperature while approaching $T_c$ from below.
Considering an increasing nature of $\sigma_{el}$ with temperature above $T_c$ reported in a number of works in the QGP
sector~\cite{Greco,Greiner-econd,Greco-econd,Cassing1,Cassing2,Qin}, it can be inferred that electrical conductivity exhibits a minimum near the QGP-hadron
cross over making it a crucial signature of the phase transition. Our results are seen to agree with most of the 
leading estimations of $\sigma_{el}$ below $T_c$. However, at low temperature our result shows 
quantitative disagreement with lattice as well as holographic approaches.

As already mentioned in the introduction, both thermal and electrical conductivities are essential inputs for the electromagnetic
particle production. Their emission rates being dependent upon the conductivities, the particle spectra and 
collective flows become sensitive to their temperature dependence as well. In light of the current situation
a rigorous study of both the quantities at finite temperature including more hadronic states and appropriate
resonances could be a topic of future investigation. Effects of finite baryonic chemical potential may be introduced through the inclusion of nucleonic degrees of freedom. In a multi-component analysis, the medium dependence of $\sigma_{el}$ incorporating the broadening of the $\Delta$ resonance and the corresponding suppression in $\pi-N$ cross-section seen in~\cite{Ghosh1} could be 
studied along the lines of~\cite{Ghosh2}. In such a treatment the exchanged mesons and baryons should also be accounted for in the electric current. More importantly there is indeed scope for improvement in the treatment of the collision term. To begin with, the RTA could be replaced with a collision term linearized e.g. using a polynomial expansion as in~\cite{Degroot}.

\section*{Acknowledgements}

Sukanya Mitra sincerely acknowledges IIT Gandhinagar, India for the institute postdoctoral fellowship. 
Snigdha Ghosh acknowledges Center for Nuclear Theory, Variable Energy Cyclotron Centre and Department of Atomic Energy, Government of India for support.

\appendix

\end{document}